\documentclass[11pt]{article}
 \usepackage{graphics}
 \usepackage{graphicx}
 \usepackage{epsfig}

\begin{document}

UDC 532.517:537.584 \hfill \vskip 1cm

\begin{center}{THE SUBGRID PROBLEM OF THE THERMAL
CONVECTION \\ IN THE EARTH'S LIQUID CORE}
\end{center}

\begin{center}{M.~Reshetnyak$^{1\,2}$, B.~Steffen$^{3}$}
\end{center}

 \footnotetext{
$^1$Institute of the Physics of the Earth, Russian Acad.~Sci,
    123995 Moscow, Russian Federation;
e-mail: maxim@uipe-ras.scgis.ru \\
    $^2$Research Computing Center of Moscow State University, 119899,
Moscow, Russian Federation;
e-mail: rm@uipe.srcc.msu.su\\
$^3$Central Institute for Applied Mathematics (ZAM)
of Forshungszentrum J$\rm\ddot{u}$lich, D-52425, J$\rm\ddot{u}$lich,  Germany;
 e-mail: b.steffen@fz-juelich.de}

\begin{abstract}
\noindent The problem of the turbulent thermal convection in the
 Earth's liquid core is considered.
 Following assumptions on decreasing of the spatial scales  due to
the rapid rotation, we propose the subgrid model of the eddy
diffusivity,  which is used in  the   large-scale model. This approach makes it possible
  to model realistic regimes with small Ekman and Rossby
  numbers ($E\sim 10^{-14}$, $R_o \sim 10^{-8}$) and  a sufficiently large
Rayleigh number $R_a \sim 10^{12}$. The obtained estimate
of the averaged kinetic energy is comparable with observations.
  The  model includes rotation of the
   solid core due to the viscous torque.
\end{abstract}

\section{Introduction}
Convection in the liquid core of the Earth, caused by the
radiactive heating and compositional processes \cite{BRo}, is the subject of
numerous researches, usually concerned with the geomagnetic field
generation, also. The last few decades saw a  fascinating development in this
area  \cite{Jones}. Based on the MHD large-scale  equations,  numerical
simulations can reproduce  different geomagnetic and geophysical
phenomena: various  properties of the
geomagnetic field (e.g., its reversals and spectrum),
eastward rotation of the inner core of the Earth as well as the
realistic ratio of the kinetic and magnetic energies
 \cite{GR95b}.

However, the wide range of spatial and temporal scales make the
direct numerical simulations (DNS) very cumbersome.
The difficulty is caused by the small values of the transport
coefficients: for instance the kinematic viscosity of the liquid core is
$\nu^M=10^{-6}{\rm m}^2$s$^{-1}$,  and the thermal diffusivity:
$\kappa^M=10^{-5}\ {\rm m}^2$s$^{-1}$
(here the subscript $^M$ corresponds to the molecular values). This gives
estimates   of the  molecular Reynolds and Peclet numbers of
$Re^M={V_{wd} L\over \nu^M}\sim 10^9$ and
$Pe^M={V_{wd} L\over \kappa^M}\sim 10^8$, where
$V_{wd}=0.2^o$ year$^{-1}$  is the west drift velocity and $L=3\cdot 10^6$\,m
is the scale of the liquid core \cite{Gubbins}, which corresponds
to the regime of the highly  developed turbulence.
In the case of the Kolmogorov's turbulence 3D DNS, simulations require
$\sim Re^{9/4}=10^{20}$ grid nodes \cite{Frisch}.
Attempts to use the exact values of these parameters on the
coarse grid lead to the numerical instabilities.
The first intuitive models in geodynamo theory
which suppressed instabilities at the small scales, e.g., the model
of hyperdiffusivity \cite{GR95b}, gave rise to new questions
concerned with interpretation of the results obtained \cite{ZH}.
The more consequential way is  an application of the
semiempirical  models of turbulence \cite{semi}. Usually, these models
are based on assumptions on the cascade transfer similar to
Kolmogorov's, which  give descriptions of  the average effect of
the small-scale field fluctuations
onto the large-scale flow  in terms of the eddy diffusivity.
The recent studies of the subgrid  \cite{Buffett} and complex models
\cite{FRS, Fricka} of the thermal convection and dynamo problems in the
rotating sphere   revealed  the  principal possibility of describing
the small-scale fluctuations in the
turbulence with the desired  Reynolds and Peclet numbers much like
Kolmogorov's  model.

 These models work up to the regime of moderate rotation speed.
  Further increase of the Coriolis
  force can  reduce the total kinetic energy and even suppress
  convection at all. From linear analysis it follows that the
  critical Rayleigh number depends on the Ekman number like
   $R_a^{cr}\sim E^{-1/3}$ \cite{Roberts68}. Even though the  molecular
  estimate of the Rayleigh number gives huge numbers
  $R_a^M\sim 10^{14}$ \cite{Kono, Gubb}, this value
  is only $5\cdot 10^2$ times larger then the critical value
  $R_a^{cr}$ \cite{Jones}. Due to the rapid  rotation of the  Earth,
the situation in the liquid core is more complicated and
assumptions on  similarity of the spectral characteristics of the fields  must
be checked very carefully. We show that the direct applications  of the
 traditional models of the turbulence, based on the mix-length assumptions,
  lead to results that differ from the observations by orders of magnitude.
The cause of such   disagreement lies in the  daily rapid rotation of
the Earth, which gives rise to new
characteristic    spatial scales in the core \cite{Meytl}.
 As a result, the energy distribution in the spectrum changes,
 which makes the application of  the Prandtl-Kolmogorov's approach
 to  the eddy diffusivity estimate difficult.
   Convection at these new scales  plays the crucial role in the
energy balance of the
 whole system and changes the estimate of the total  energy by orders of
 magnitude. Even a simple  account of these  effects
 leads to  essential change of the rate of the energy dissipation and thus
 to a better agreement of LSS models with the observations.

 In the section \ref{An}  we introduce the large-scale equations
 of the thermal convection and consider the Prandtl-Kolmogorov's
 assumptions  on the eddy diffusion. In the section \ref{Rol} we recall
the basics of
  convection in a rapidly rotating body and estimate the subgrid
  diffusion. Afterwards, this  estimate  is used in the
  large-scale model, section \ref{Bas}.  The discussion of results   is
in section \ref{Disc}.

\section{The large-scale equations}\label{An}
The problem of the thermal convection in the Earth's core can be
reduced to the problem in the spherical shell.
Let the surface of the sphere, radius $r_0$ (in the
spherical system of coordinates $(r,\,\theta,\, \varphi)$),
rotating with angular
velocity $\Omega$ around the $z$-axis. This sphere contains a
concentric solid inner sphere, radius $r_i$, and the outer
spherical layer ($r_i< r< r_0$) is filled with an incompressible
liquid (${\bf V} = 0$). The inner sphere is allowed to
rotate freely around the $z$-axis due to viscous torque.
Convection in the Boussinesq approximation in the outer sphere is
described by the Navier-Stokes equation and by the heat flux
equation.
Choosing $L = r_0$ as the unit of length, velocity ${\bf  V}$,
time $t$ and pressure $p$ can be measured in units of
$\kappa^M/L,\, L^2/\kappa^M$ and $2{\Omega}\rho\kappa^M$,
respectively.  Then, the governing
equations can be written in the form
\begin{equation}
R_o^M\left[ {\partial {\bf V}\over\partial t}+ \left({\bf V}\cdot
\nabla\right) {\bf V}\right] = -\nabla p + -{\bf {1}_z}\times{\bf V} +
R_a^M Tr{\bf{1}_r}+
E^M\nabla\cdot \buildrel\leftrightarrow\over {{\bf S}},
\label{Nav}
\end{equation}
\begin{equation}
{\partial T\over\partial t}+\left({\bf V}\cdot\nabla\right)
\left(T+T_0\right)= \nabla\cdot( \nabla T),\label{therm}
\end{equation}
where
$\bf 1_z$ is the unit vector in $z$-direction,
 $\buildrel\leftrightarrow\over{\bf S}$ is the rate of the
strain tensor and $T$ the temperature fluctuations relatively to the
imposed profile
$T_0={r_i/r-1\over 1-r_i}$.
 The molecular Rossby $R_o^M$, Ekman $E^M$ and Rayleigh $R_a^M$
numbers appear in the equations
\begin{equation}\begin{array}{l}
R_o^M={\kappa^M\over 2{\Omega}L^2},\qquad\qquad\quad
E^M={\nu^M\over 2{\Omega}L^2},\qquad R_a^M={\alpha g_0\delta T L\over
2{\Omega}\kappa^M},
 \end{array}
\label{Ra_def}
\end{equation}
 where $\alpha$ is the coefficient of thermal
expansion, $g$ the gravity acceleration and $\delta T$
is a temperature unit ($\delta T\sim 10^{-4}$K, see \cite{Jones}).
 It should be mentioned that the Rayleigh number for
non-rotating bodies is usually given in the form
$\widetilde{R}_a^M=\alpha g\delta T L^3/\kappa^M\nu^M$   and that
$R_a^M=E^M\widetilde{R}^M_a$.

The solid inner
sphere is allowed to rotate freely around the $z$-axis due to
viscous torque. The dimensionless momentum equation for the
angular velocity $\omega$ of the inner sphere ($0 <r < r_i$) has
the form
\begin{equation} R_o^MI{\partial\omega\over\partial t} =
 r_i E^M\oint\limits_{\cal S}S_{\varphi r}\big|_{r=r_i}
 \sin\vartheta \, d{\cal S},\label{core0}
\end{equation}
where $I$ is the moment of inertia of the inner sphere $\cal S$ and
$S_{\varphi r}$ is a component of the strain tensor in the
  spherical system of coordinates \cite{Landau}.
 Equations (\ref{Nav}--\ref{therm}, \ref{core0})  are accompanied by
the non-penetrating and no-slip boundary condition for velocity $\bf V$
and zero temperature fluctuations at the  shell boundaries.

The system (\ref{Nav}-\ref{therm},\ref{core0}) was successfully  studied in
the regimes of the laminar convection
using different numerical approaches \cite{Christensen, Jones}.
 However, these regimes are still very far  from the  desired
 estimates for the Earth's liquid core
$R_o^M= 10^{-8}$, $E^M= 10^{-14}$ and $R_a^M= 10^{14}$ \cite{Jones}.
 Attempts to approach to these  parameters using DNS  caused
numerical instabilities and required application of turbulencs
models \cite{Buffett}. However, even the direct usage of the known
models of turbulence is not trivial.

  To support  this point we offer a simple estimate of the eddy
 diffusivity, based on the most popular mix-length model of
 the  turbulence.
 Following the Prandtl-Kolmogorov hypotheses, the eddy diffusion at
the scale $l$  can be estimated as
  $\nu^T=(\varepsilon l^4)^{1/3}$, where $\varepsilon=v^3/l$ is
   the  rate of energy dissipation and $v$ is a velocity at the
scale $l$. Even the largest estimate, based
  on the main scale $l=L$ and the west drift velocity
$V=3\cdot 10^{-3}$m\,s$^{-1}$,
    gives $\nu^T= 2\cdot 10^3$m$^2$s$^{-1}$, giving an
   Ekman number  of order $E^T=2\cdot 10^{-5}$. The more
   realistic estimate with
   $V_l=\delta V\sim V ({{\it l}\over L})^{1/3}$
   and the usual grid scale of ${\it l}\sim 3\cdot 10^{-2}L$ gives
   $\nu^T= 15$ and $E^T=10^{-7}$. On the other hand,
  this estimate of $E^T$ would require resolution of about
   $N_\varphi\sim 2\cdot 10^2$
   columns  \cite{Roberts68},  which need  use of the most powerful
modern computers.
    All this  means that this estimate of $\nu^T$ will  not
    provide the smooth field behaviour of fields assumed
    in the Kolmogorov's turbulence, when $E^T \ge 1$.
    Thus, the traditional methods underestimate the eddy diffusion
    $\nu^T$.

    Such situation corresponds to the case, where the classical
    ideas on the direct cascade of energy from the main scale $L$
    to the dissipative scale  are violated and additional information
    on   $\varepsilon$ at the dissipative
      scale is needed. As we see below in the section \ref{Rol},
      it appears that  in the case of the rapid rotating body the  energy
      in the spectrum  is
      shifted  to the small scales, those which  DNS cannot
      resolve even at the onset of convection.
      This is the reason, why any attempts to
     estimate $\nu^T$ in the turbulent regime at scales compared with the grid
     resolution, lead to the non-selfconsistent behaviour of the
     turbulent model.

The way out of such difficulties is  make  proper assumptions on the
spectral properties of the solution in the range of the high wave numbers.

\section{The model of columns}\label{Rol}
 The origin of the problem can be seen from the analysis of
 the linerized  system
(\ref{Nav}-\ref{therm}) at the onset of convection in the limit of
small Rossby and Ekman numbers.
 As it was shown in \cite{Roberts68} already (see
also recent paper \cite{Mussa}), at the onset of convection
the structure of the flow tends to develop
columns along  $z$-direction, such that
$\partial / \partial \varphi\sim {\cal O}(E^{-1/3})$,
$\partial / \partial s\sim {\cal O}(E^{-1/6})$,
$\partial / \partial z\sim {\cal O}(1)$,
 when $E=R_o\to 0$.
 Linerization of the system
 (\ref{Nav}-\ref{therm}) leads to the balance of the Archemedean and
viscous terms in  the Navier-Stokes equation:
  $R_a\, T\sim E^{-1/3}V$. The balance of the convective and viscous terms
in the   heat-flux equation gives $V\sim E^{-2/3} T$,
 from   which follows the estimate of the critical Rayleigh
number $R_a^{\rm cr}\sim E^{-1/3}$.
(For convenience we omitted index $^M$.)
Such, at the onset of convection for system (\ref{Nav}-\ref{therm}), the
flow is anisotropical with the smallest scale $l_E\sim E^{1/3}L$,
defined by  the balance of the Coriolis and viscous forces.
Note that the scale $l_E\sim 10^{-5}$
 is beyond the level  of DNS.  If this asymptotic is correct, the
 critical Rayleigh number in the Earth's core is
$R_a^{cr}\sim 10^5$ \cite{Jones}.
  As we show below, the
predicted  column-like form of the flow is  very important for estimates of
the subgrid  dissipation in the liquid core.

The main  assumption is, that  even in  the turbulent regime believed to be in
the Earth's liquid core, the flow tends to elongated structures with
 the smallest scale  $l_E$,  predicted by the linear analysis.
  It is from this scale ideas of the direct cascade of energy are
  applicable. To simplify the problem, we estimate the isotropical eddy
  diffusion,
  based on the  scale ($l_E$).
In particular, instead of the
    estimate of velocity gradient at the subgrid scale $l$:
      $V'\sim  \delta V/l$,  we use  $V'\sim E^{-1/3} \delta V$,
      where $\delta V\sim 0.3 V$ is the average variation
       of velocity at the scale  $l$. In this case the
  estimate of the eddy diffusion gives
  $\nu^T\sim l^2 V'\approx 5\cdot 10^4$m$^2$\,s$^{-1}$ and $E^T\sim
  5\cdot 10^{-4}$. This estimate of the turbulent Ekman number
  corresponds to $N_\varphi\sim 10$ columns
   which can be resolved in the large-scale models with the desired
  accuracy. To demonstrate these arguments, we propose a
  simulations of the system (\ref{Nav}--\ref{therm}, \ref{core0}) with
  the given eddy  diffusion $\nu^T$ estimated as above.

\section{Turbulent model. Results of calculations.}\label{Bas}
\begin{figure}\vskip -1cm \hskip 5cm
\epsfxsize=30cm\epsffile[150 18 1200 700]{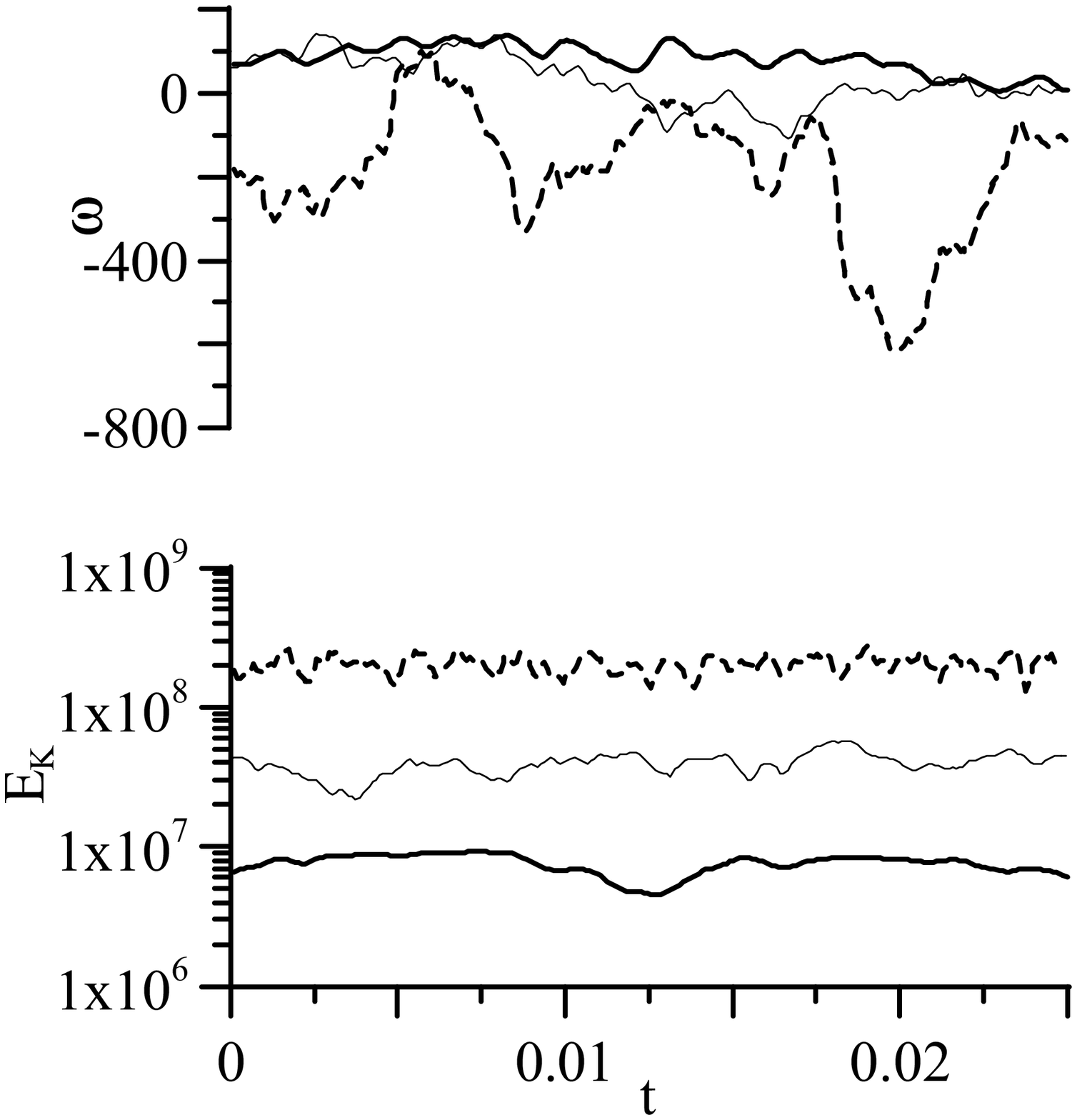}\vskip -3cm
\caption{Evolution of the angular velocity  of the liquid core
$\omega$
 and averaged over the volume kinetic energy $E_k$ for
 $R_o=4\cdot 10^{-7}$, $E=10^{-x}$;
 $R_a^T=10^6$ -- thick line,
 (2) -- $R_a^T=10^7$ -- thin line,
 (3) -- $R_a^T=10^8$ -- dashed line.} \label{fig1}
\end{figure}

Equations (\ref{Nav}--\ref{therm}, \ref{core0}) are solved using the
control-volume method (Simple algorithm)
\cite{Patankar} on the staggered grid
($n_{\rm r},\, n_\theta,\, n_\varphi)=(45,45,64)$.
 This method is based on the finite-difference
 approximation and demonstrates very high
numerical stability for the regimes with strong
convection\footnote{See also some special questions of the
control-volume method for the full dynamo problem in the  sphere
in \cite{HR03}.}.
 For ease of calculation, we renormalize equations
  (\ref{Nav}--\ref{therm}, \ref{core0})  using
 turbulent diffusion units, so that instead of $\kappa^M$ the
 $\widehat{\kappa}=1$m$^2$\,s$^{-1}$ was used.
 Then, the dimensionless parameters are:
  $R_o^T={\widehat{\kappa}\over 2\Omega L^2}=4\cdot 10^{-2},$
 $E^T={\nu^T\over 2\Omega L^2}=10^{-3}$.
  We consider three regimes with turbulent Rayleigh numbers
 $R_a^T={\alpha g_0\delta T L\over 2{\Omega}\widehat{\kappa}}=
 10^6$, $10^7$, $10^8$ (see   the time evolution of the kinetic
 energy $E_K$ in Fig.~\ref{fig1}).
The corresponding
  Reynolds numbers averaged over the shell volume
$Re^M={\widehat{\kappa}\over \nu^M}\sqrt{2
E_K}$ are $3\cdot 10^9$, $6\cdot 10^9$ and $2\cdot 10^{10}$, c.~f.
with the molecular Reynolds number for the Earth's core based on
the west drift velocity $R_e^{Earth}\sim 10^{9}$.

\begin{figure}\vskip -10cm \hskip -6cm
\epsfxsize=37cm \centering\epsffile[5 50 1200 1200]{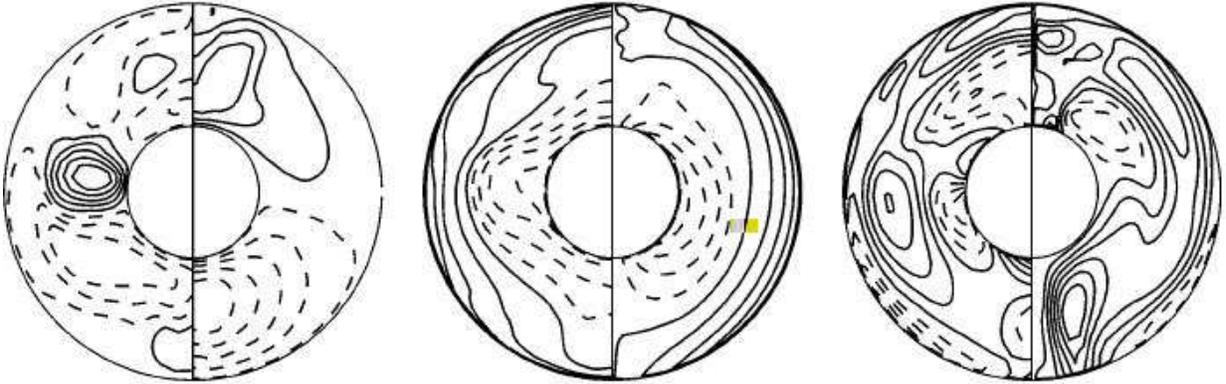}
\vskip -8cm
\caption{The snapshots of the  velocity field components
$(v_r,\,v_\vartheta,\,v_\varphi)$ (from left to right) for the equatorial sections
(the left half of the plane):
$(-700,\,1200)$,
$(-4700,\,4100)$,
$(-1200,\,1800)$
and meridional sections for axi-symmetrical parts of the
fields (the right half):
$(-2400,\,1800)$,
$(-3800,\,3100)$,
$(-400,\,1000)$.
Numbers in round brackets indicate ranges.} \label{fig2}
\end{figure}

Characteristic  snapshots of the large-scale velocity
$r,\,\theta,\,\varphi$-components are
presented in Fig.~\ref{fig2}. The observed curls in $r,\, \varphi$-projections
 corresponds to the columns parallel to $z$-axis. These columns may
drift in the $\varphi$-direction.
In its turn, the non-zero viscous gradient
$\nu {\partial \over \partial r}\left(V_\varphi\over r\right)_{r={r_i}}$
  causes rotation
 of the inner core, (see evolution of the angular velocity of the inner core
  $\omega$ in Fig.\ref{fig1}).
  Here the positive value of $\omega$ corresponds to the
   eastward direction, known to occur in the Earth \cite{Song96}.
    We emphasize that  these maps are  a
product of averaging of the small-scale
($l_E\sim{\cal O}(E^{M\, 1/3})=10^{-5}$) structures.
So far, the micro-scale Reynolds number $r_e$ at the scale $l_E$  is still
larger then unity, and  the inertial  spectrum for the scales
 smaller then $l_E$ exists.  An estimate of $r_e={vl\over \nu^M}$
 with $l=E^{M\, 1/3}$ and $v=0.1V$ gives $r_e\sim 10^3$.
   This spectrum has two parts with the transition point defined by the
   balance of the inertial and Coriolis terms:
   $l_\Omega\sim R_o v $.
 The turbulence in the range of $l_E\le  l_\Omega$ is influenced
     by rotation and  the kinetic energy
  spectrum is $E_l\sim l^2$  \cite{Zhou}. For
   the scales smaller then $l_\Omega$ up to the dissipative scale
   $l_d=R_e^{-3/4}$ the Kolmogorov's spectrum
    $E_l\sim l^{5/3}$ reappears.

Summarizing the obtained results we conclude, that based  on the
realistic values of the Rossby and Rayleigh numbers and on
assumptions on the spectrum of the flow in the liquid core we
obtained a value of the  kinetic energy $E_K$ comparable with the
observations. Having in mind that the velocity field and the eddy
diffusion are connected in our model, we consider this agreement
to be worth to note.

\section{Conclusions}\label{Disc}
We propose the scenario of the turbulent thermal convection in the rapid
rotating body, when the Coriolis force shifts the system to the origin
of the small scales already at the onset of convection, and show  that
  further increasing of the
 intensity of the heat sources leads to a turbulent regime, which
 is still far from the Kolmogorov's case. It appears that
 predictions of the linear analysis at the onset of convection are
 applicable to  the eddy diffusion estimate in the regime of the fully
 developed turbulence. Though the original problem is highly
 anisotropical, the  ``isotropical'' estimate  of
 the eddy diffusion gives a kinetic energy of the system comparable
with the observations.
  Note that introduction of the magnetic field will
 not change the problem in principal, because at the scales $l_E$
 considered the corresponding micro-scale magnetic Reynolds number
 is already $r_m\ll 1$ and the magnetic field decays due to the Ohmic
 dissipation process.  On the other hand, it is not yet clear how the
west drift velocity relates to the flow at the scales $l_E$ and different
 interpretations of observations can exist. This question requires the
 solution of the full dynamo  problem.

\section*{Acknowledgements}
RM  is grateful to Central Institute for Applied Mathematics (ZAM)
of Forshungszentrum in J$\rm\ddot{u}$ulich for hospitality. This
work was supported by
 the
Russian Foundation of Basic Research (grant 03-05-64074).



\end{document}